# Reducing Latency for Multimedia Broadcast Services Over Mobile Networks

Carlos M. Lentisco, Luis Bellido, and Encarna Pastor

*Abstract*—Multimedia services over mobile networks pose several challenges, such as the efficient management of radio resources or the latency induced by network delays and buffering requirements on the multimedia players. In Long Term Evolution (LTE) networks, the definition of multimedia broadcast services over a common radio channel addresses the shortage of radio resources but introduces the problem of network error recovery. In order to address network errors on LTE multimedia broadcast services, the current standards propose the combined use of forward error correction and unicast recovery techniques at the application level. This paper shows how to efficiently synchronize the broadcasting server and the multimedia players and how to reduce service latency by limiting the multimedia player buffer length. This is accomplished by analyzing the relation between different parameters of the LTE multimedia broadcast service, the multimedia player buffer length, and service interruptions. A case study is simulated to confirm how the quality of the multimedia service is improved by applying our proposals.

## I. INTRODUCTION

NOWADAYS, the widespread use of smartphones, as well as the availability of higher bandwidths in next generation mobile networks, have led to a growing demand of live video streaming services that users can access through multimedia players running on their mobile devices. In this context, quality of service can be affected by both service interruptions due to mobile network transmission errors, and the latency introduced by error recovery mechanisms.

The 3rd Generation Partnership Project (3GPP) defines how to broadcast multimedia content over Long Term Evolution (LTE) using Dynamic Adaptive Streaming over HTTP (DASH) technologies [1]. DASH, originally defined to work over a unicast bi-directional connection, provides a standard to split a multimedia stream into segments (files) of the same duration that can be retrieved from a web server. A Media Presentation Description (MPD) file includes all the parameters needed by a multimedia player to access these segments and play the multimedia stream. DASH defines also how to provide different representations of the multimedia content, coded with different qualities, so a multimedia player can select a representation that best fits both the characteristics of the mobile device and an estimation of the available bandwidth.

With the aim of better managing the available radio resources, the 3GPP proposes to combine DASH with a multicast delivery of multimedia segments to the mobile devices, using the evolved Multimedia Broadcast and Multicast Service (eMBMS) [2]. In this case, a single representation of the multimedia content is pushed to the devices through the eMBMS channel. So the adaptability of DASH is not used because the receiver is not given the choice of different representations. Since the transmission over eMBMS can suffer from errors, the 3GPP has proposed the use of Application Layer Forward Error Correction (AL-FEC) [3] and unicast recovery techniques. The multimedia segments being pushed over eMBMS are protected by an AL-FEC and the receiving device will use the AL-FEC to try to recover the segments that are affected by transmission errors. The segments that cannot be recovered by the AL-FEC are retrieved by the multimedia player using HTTP, as defined by DASH. But the DASH player does not select a representation based on device characteristics or available bandwidth. Instead, it either plays the representation that it receives over eMBMS, or it requests that same representation from a web server, when transmission errors cannot be corrected by the AL-FEC.

In this paper, we show how this combination of AL-FEC and HTTP error recovery can lead to synchronization problems between the broadcasting server and the multimedia players. As a result, there would be unnecessary transmissions over HTTP for data that is being sent over eMBMS. By analyzing the different components of the latency, we propose a solution to avoid these unnecessary transmissions.

We will also analyze in more detail one of the components of the latency: the buffering delay at the multimedia player. A long enough buffer avoids service interruptions when the AL-FEC fails and a segment needs to be recovered by HTTP.

This research has been partially supported by the Spanish Ministry of Economy and Competitiveness in the context of the project GREDOS ref. TEC2015-67834-R (MINECO/FEDER, UE).

Carlos M. Lentisco is with the Dep. of Telematic Engineering, Universidad Politécnica de Madrid, 28040 Madrid, Spain (e-mail: clentisco@dit.upm.es).
Luis Bellido is with the Dep. of Telematic Engineering, Universidad Politécnica de Madrid, 28040 Madrid, Spain (e-mail: lbellido@dit.upm.es).
Encarna Pastor is with the Dep. of Telematic Engineering, Universidad Politécnica de Madrid, 28040 Madrid, Spain (e-mail: epastor@dit.upm.es).

But an over-provisioning of the buffer used by the multimedia players increases the end-to-end latency. We analyze the relation between different parameters of the LTE multimedia broadcast service, the multimedia player buffer length and the service interruptions, to show how to reduce service latency by limiting the multimedia player buffer length. This can be particularly relevant for live streaming services, that would benefit from having the minimum buffering delay that guarantees service continuity.

Section II provides a brief description on the architecture proposed by the 3GPP to support multicast streaming services.

Section III presents a delay analysis of an LTE multimedia broadcast service that allows the calculation of the earliest availability time for a DASH segment that is delivered over eMBMS. It also shows how this value can be used in DASH to synchronize the unicast error recovery requests initiated by the multimedia player with the delivery over eMBMS.

In Section IV, the buffer requirements of the multimedia player are analyzed, taking into account the recovery of segments via HTTP. Using the packet error rate and the available bandwidth as the main parameters, Section IV shows how to calculate the minimum buffer level needed to avoid buffer starvation, so it is possible to avoid service disruptions with a minimum latency.

Section V discusses how it is possible to further reduce latency by selecting an adequate duration for the multimedia segments, taking into account the limitations due to the use of AL-FEC techniques and the recovery of lost segments through HTTP. Finally, Sections VI and VII discuss the related work and provide the conclusions of this work.

## II. HYBRID FLUTE/DASH ARCHITECTURE

Fig. 1 shows the architecture proposed by the 3GPP to support LTE multimedia broadcast services. DASH, initially designed to work as a multimedia streaming solution over HTTP, has been adapted for multicast streaming services over LTE using the eMBMS channel. However, even if the use of eMBMS allows sending the same multimedia segment to multiple receivers using a common channel, the eMBMS channel is unidirectional and unreliable, so recovery mechanisms must be used to ensure a correct delivery of the DASH segments over eMBMS.

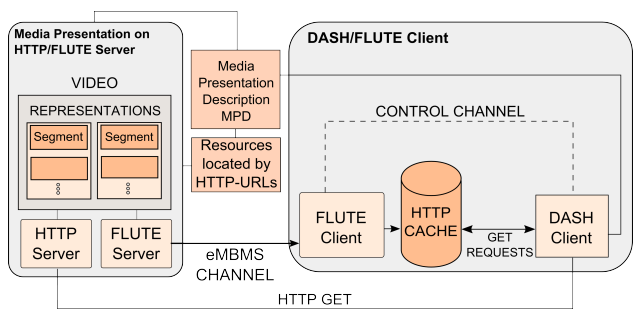

Fig. 1. Hybrid FLUTE/DASH architecture. The figure shows the components involved in the delivery of DASH segments over eMBMS in combination with unicast error recovery using HTTP.

The 3GPP has proposed both application and physical level mechanisms to overcome the lack of reliability in eMBMS.

At the physical level, the 3GPP proposes the use of Physical Layer Forward Error Correction (PHY-FEC) techniques to protect data against errors during its transmission over the wireless LTE channel. Errors are also mitigated by the use of Multicast/Broadcast over Single Frequency Network (MBSFN). MBSFN is a transmission mode that improves the Signal to Interference plus Noise Ratio (SINR) of the multicast receivers by defining how several base stations cooperate to transmit the same signal with very precise time/frequency synchronization.

However, the use of MBSFN and PHY-FEC techniques is not enough to provide a reliable delivery of multimedia segments (files) over LTE. This is the reason why these techniques at the physical level are combined with two additional mechanisms at the application level, AL-FEC and unicast error recovery. AL-FEC techniques consist of the generation of redundancy data that are sent together with the file to be transmitted. A receiver is able to recover a file if enough data have been received, either from redundancy or from the original object. Otherwise, the data is discarded and the file cannot be delivered to the application.

A multimedia content encoded as defined by DASH is composed of several segments. Each segment is independently protected using the AL-FEC and then it is sent to the multiple receivers using eMBMS. For the delivery of DASH segments over eMBMS, the 3GPP proposes the use of the File Delivery over Unidirectional Transport (FLUTE) protocol [4]. FLUTE is a protocol particularly suited for the unidirectional delivery of files in multicast networks, since it works over UDP and it can be used together with AL-FEC techniques.

However, the AL-FEC decoder might not be able to recover a multimedia segment when there is a high error rate in the eMBMS channel. In this case, the 3GPP proposes to recover the lost segments using HTTP. The whole system works as follows (Fig. 1). Using FLUTE, the DASH multimedia segments are pushed over eMBMS to the multiple receivers connected to the eMBMS channel. When a FLUTE client receives a DASH segment, it copies it to a local HTTP cache. A DASH client request goes first through a local HTTP cache, so a segment that is received correctly over FLUTE is delivered from the cache to the DASH client. In case a segment is not available in the cache, it is retrieved from the content server via HTTP using the regular unicast access channel.

## III. DELAY ANALYSIS OF A LIVE STREAMING SERVICE OVER THE LTE BROADCAST ARCHITECTURE

In an LTE multimedia broadcast service, a multimedia segment suffers a series of different delays from its generation in the content server until its playback in the multimedia player, as shown in Fig. 2. Each of these different delay components is described in the following subsections. Table I is provided as a reference for the variables and nomenclature used.

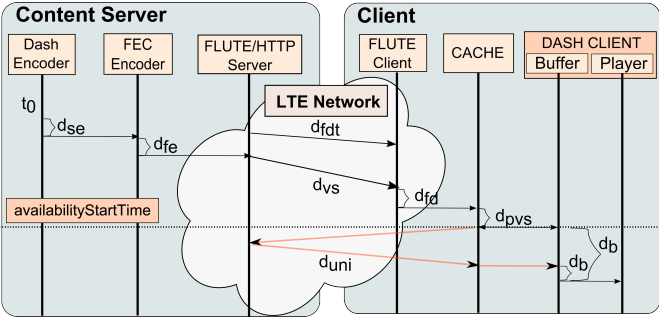

Fig. 2. Delay analysis. This figure shows the delays suffered by a DASH segment that is sent using eMBMS and the earliest availability time of the segment on the mobile device.

TABLE I
DEFINITION OF VARIABLES AND NOMENCLATURE

| Name | Description |
|---|---|
| $CR$ | AL-FEC code rate. It is defined as the ratio between the $k$ source symbols and the $k+r$ symbols |
| $d_b$ | Buffer level in seconds |
| $d_{fd}$ | AL-FEC decoding delay for a multimedia segment |
| $d_{fdt}$ | Transmission delay of an instance of the FDT |
| $d_{fe}$ | AL-FEC encoding delay for a multimedia segment |
| $d_{pvs}$ | Margin of security used to avoid synchronization problems |
| $d_{se}$ | Multimedia segment generation delay |
| $d_{uni}$ | Multimedia segment unicast recovery delay |
| $d_{vs}$ | Multimedia segment transmission delay over FLUTE |
| $D_T$ | Multimedia segment transmission delay over unicast |
| $ISD$ | Inter-Site Distance |
| $k$ | Number of symbols composing an AL-FEC source block |
| $MCS$ | Modulation and Coding Scheme |
| $n$ | Number of symbols received by the AL-FEC decoder |
| $P(f_{RC})$ | Probability of the FEC decoder not recovering a segment due to network errors |
| $P(f_{RC})^m$ | Probability of losing $m$ video segments |
| $PER$ | Packet Error Rate |
| $r$ | Number of repair symbols generated by the AL-FEC encoder |
| $SDR$ | Service Data Rate. Effective bandwidth for delivering a video over eMBMS |
| $T_{Seg}$ | Duration of a multimedia segment |

## A. Generation of DASH Video Segments

The delay analysis starts at instant $t_0$, representing the instant the first byte of a live video signal is received by the content server from the source. In order to prepare the video data to be sent over the eMBMS channel, a DASH encoder located in the content server is in charge of encoding the input stream and splitting it into segments of a predefined duration. Often, it is necessary to convert the video, applying a video compression format such as H.264. In the particular case of an LTE multimedia broadcast service, only one representation is needed to be sent over the multicast channel, so video segments are coded with a single bitrate. In conclusion, both segmentation and coding are needed to generate a DASH video segment. In Fig. 2, the time taken to generate a DASH video segment is denoted as $d_{se}$.

## B. Protecting Video Segments Using AL-FEC Techniques

At present, the AL-FEC technique standardized by the 3GPP to provide a reliable transmission over eMBMS is Raptor [5]. Raptor is a block code that allows splitting a data stream into several files, also known as source blocks, so each source block can be independently coded. In an LTE multimedia broadcast service, each multimedia segment generated is a different source block.

A Raptor encoder divides each source block into $k$ fragments of the same size, called source symbols. By using these symbols, the encoder generates the redundancy data, which consists of $r$ symbols also of the same size (these symbols are known as repair symbols). A metric that is normally used to express the amount of redundancy introduced by the transmitter is the AL-FEC code rate, which is defined as the ratio between the $k$ source symbols and the $k+r$ symbols resulting of the AL-FEC encoding process. In this case, these $k+r$ symbols representing the multimedia segment and the redundancy data, are sent over the eMBMS channel. The decoder is able to retrieve the multimedia segment from any set of encoding symbols only slightly more in number than the number of source symbols. The performance of the decoding process is greatly improved by the use of larger source blocks [3]. However, the use of multimedia segments of a long duration can increase the delay (Section V), and hence, reduce the quality of service perceived by the users.

In Fig. 2, the time taken to encode a video segment together with the resulting AL-FEC code is denoted as $d_{fe}$, whereas the time needed to decode a video segment is denoted as $d_{fd}$.

## C. Delivery of Live Video Using FLUTE

In a multicast transmission over eMBMS, video segments are delivered within a FLUTE session. Before receiving a file, a FLUTE receiver needs to receive the attributes of the FLUTE session and a description of the properties of the file [4]. This information is sent in-band within the file delivery session through an XML file called File Delivery Table (FDT).

In a live streaming service, the video segments are progressively generated, so it is necessary to notify, also progressively, the properties of each of the video segments that are sent using FLUTE over the multicast channel. An initial approach, consisting of sending an instance of the FDT describing the properties of all the video segments, is not possible, because one of the properties that needs to be included is the segment size, which is unknown beforehand [4]. Instead, multiple instances of the FDT are sent, each one describing the properties of the video segment that is already available to be sent. For this analysis, the delay of generating each FDT instance is not considered, as they are small files that only include the metadata of one video segment.

If an AL-FEC technique is used, the FEC Object Transmission Information [6] must be sent within the FLUTE session [4]. This paper considers using the FDT to transmit the FEC Object Transmission Information, because it requires a lower overhead compared to sending this information in-band. This information depends on the FEC schema used. In the case of Raptor, this is the information needed by the receivers to decode a block of data, i.e., the length of the video segment and the symbol size. Therefore, it is possible to send an instance of the FDT describing a new video segment while the FEC for the segment is being coded using Raptor.

Finally, the time needed to deliver an object over the eMBMS channel can be calculated based on two parameters: object size and link transmission rate. The link transmission rate for the eMBMS session is fixed because radio resources are pre-allocated for this service.

With a fixed transmission rate, the time taken to transmit an instance of the FDT ($d_{fdt}$) is considerable shorter than the time of transmitting a video segment ($d_{vs}$), so that $d_{fdt}$ can be ignored. Note that the FDT size can be a few hundreds of bytes, whereas the size of a video segment is larger by more than one order of magnitude.

### D. Protection Against Delay Variations

The DASH standard defines a mechanism to inform clients about the availability of a new segment in the server. This mechanism is based on the use of the *availabilityStartTime* field in the MPD. The standard defines *availabilityStartTime* as the earliest availability time (in UTC) for any segment in the multimedia presentation [1]. Lohmar *et al.* [7] analyze the impact of that value on a DASH unicast transmission, taking into account the possible variation of different delays. For example, when a Variable Bitrate Coding (VBR) scheme is used to code a video fragment, the time taken to code a video segment is not constant. A possible consequence is that clients poll the availability of a video segment some time before it is really available. A frequent poll can overload the server, so Lohmar *et al.* [7] propose to delay the HTTP requests by increasing *availabilityStartTime* taking into account possible variations in delay.

In the case of an LTE multimedia broadcast service, the impact of the *availabilityStartTime* is different, but a similar margin of security is needed. The time taken for a video segment to be pushed to the local cache of the clients depends on the delay components previously analyzed, which are variable. In this case, the possible consequence is that DASH clients request a video segment that is not yet available in the cache, but that will be received later via eMBMS. This is not desirable, because the video segment will then be unnecessarily requested and transmitted over HTTP, thus wasting communication resources. Therefore, for an LTE multimedia broadcast service it is also necessary to introduce a margin of security that takes into account the delay variations of the previous components. This ensures that video segments are not requested before they are copied to the receiver cache. This margin of security is denoted as $d_{pvs}$ in Fig. 2.

### E. Unicast Delivery

If a client requests a video segment that is not available in the cache, a session is established between the client and the server to deliver the segment over HTTP. Taking into account the margin of security introduced by $d_{pvs}$, this can only happen in two different scenarios:

- when a video segment has been incorrectly decoded by the FEC decoder, or,
- when the video segment has been correctly decoded, but the instance of the FDT that describes the video segment has been lost.

The use of HTTP to deliver a video segment adds an additional delay, denoted as $d_{uni}$ in Fig. 2. This delay depends on several factors. For instance, all the LTE terminals using unicast services on a cell share the available radio resources. These resources are dynamically allocated to the terminals, so it is not known for sure how much bandwidth will be allocated to each terminal. In particular, the case in which the transmission delay of the video segment is longer than the duration of the segment has a significant impact on the total delay (Section IV).

### F. Client Buffer

Buffering of video in the client is necessary to ensure a seamless playback. In unicast streaming services, the buffer level can be used by a DASH client to adapt the retrieval of segments in response to changing network conditions [8], [9]. For instance, when the buffer level falls below certain threshold, it can be interpreted as not having sufficient available bandwidth to download video segments at the requested quality, so the client can switch to another representation of the video encoded with a lower bitrate. Thus, the buffer level is used by the adaptation logic of the DASH client to avoid service disruptions.

In a multicast streaming service over LTE, buffer requirements are different. Only one representation is sent over eMBMS, so the buffer level cannot be used to switch between different representations. On the other hand, a video segment that is lost in the eMBMS channel needs to be recovered via HTTP, so a long enough buffer is needed to avoid buffer starvation during the recovery of lost segments. The buffer level that is required to start the playback of a video is specified by the *minBufferTime* field of the MPD. If a low value of *minBufferTime* is set, and there is a high error rate that implies the loss of several video segments in the multicast channel, buffer starvation can happen. To solve this problem a strategy to calculate the minimum buffer level that is needed to avoid possible video freezes during the playback is proposed in the next section. This buffer level will be characterized by its value in seconds, denoted as $d_b$ in Fig. 2.

### G. Playback Deadline

Based on the delay analysis presented above, one of the proposals of this paper is to set the *availabilityStartTime* field of the MPD to a value that guarantees that a video segment correctly delivered over eMBMS is available on the local cache of the clients before it is requested over HTTP. This avoids the synchronization problems that would lead to unnecessary transmissions over HTTP. Taking into account the different components of the delay analyzed above, the proposed value for the *availabilityStartTime* field can be calculated as:

$$availabilityStartTime = d_{se} + d_{fe} + d_{vs} + d_{fd} + d_{pvs} \quad (1)$$

This time is depicted in Fig. 2 as a dashed line together with the tag *availabilityStartTime*. The playback deadline, defined as the instant in which a video segment content is started to be played by the DASH client is then calculated as:

$$Playback_{deadline} = availabilityStartTime + d_b \quad (2)$$

where $d_b$ is the buffer level. In an LTE multimedia broadcast service, the buffer level determines the seconds of video that must be stored in the buffer before starting the playback. In the following section, we are going to analyze how to calculate the minimum buffer level needed to avoid service disruptions when video segments are lost in the multicast channel and the multimedia player recovers the segments using HTTP.

## IV. Dimensioning the Multimedia Player Buffer

This section shows a method to calculate the minimum buffer level that prevents buffer starvation without unnecessarily increasing the initial startup delay of the video playback. This buffer level will be calculated based on the eMBMS packet error rate, the Round Trip Time (RTT), and the downlink transmission delay for the unicast channel. By combining this method with the simulation results of different LTE broadcast scenarios, the impact of different deployment parameters on the buffer level is analyzed.

### A. Buffer Requirements of the Multimedia Player

The use of AL-FEC techniques is not always sufficient to ensure a reliable delivery of DASH video segments, so video segments lost in the multicast channel are recovered via HTTP. However, there is no reservation of bandwidth for this communication. Monserrat *et al.* [10] discuss that the maximum bandwidth allocated to a terminal decreases as the number of terminals in a cell increases. Therefore, when there are multiple terminals in a cell, the bandwidth available for the unicast video segment recovery can be lower than the bitrate of the video. In this case, a video segment lost in the multicast channel is recovered through HTTP later than expected, which could lead to buffer starvation.

To calculate the minimum buffer level needed to avoid buffer starvation, the worst possible case in term of losses is considered, i.e., a case in which a consecutive sequence of video segments are lost. The probability of losing a video segment is calculated as the probability of the FEC decoder not recovering a segment due to errors in the packets received through the eMBMS channel. This probability is denoted as $P(f_{RC})$, and is given as:

$$P(f_{RC}) = \sum_{n=0}^{k+r} P(f_{RC} \mid n) \cdot P(N = n) \quad (3)$$

where $P(f_{RC} \mid n)$ is the failure probability of the Raptor decoder in case of receiving n encoding symbols and where $P(N=n)$ is the probability of receiving $n$ symbols.

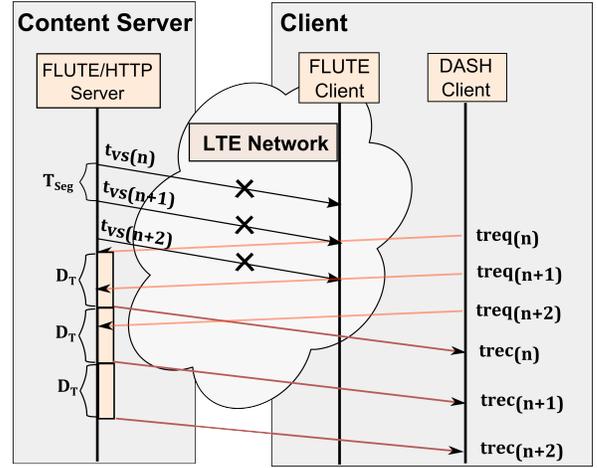

Fig. 3. Delay analysis. This figure shows the time that would be needed to recover, via HTTP, a burst of video segments lost in the eMBMS channel.

According to the mathematical model presented by Stockhammer *et al.* [11], $P(f_{RC} \mid n)$ can be calculated as:

$$P(f_{RC} \mid n) = \begin{cases} 1 & \text{if } n < k \\ 0.85 \times 0.567^{n-k} & \text{if } n \geq k \end{cases} \quad (4)$$

On the other hand, since each symbol is encapsulated in a packet, given a Packet Error Rate (PER), the probability of receiving $n$ symbols can be calculated by using a binomial distribution, given as:

$$P(N = n) = \binom{k+r}{n} \cdot (1-PER)^n \cdot PER^{k+r-n} \quad (5)$$

In order to calculate the minimum buffer level, a *threshold* is used to limit the probability of losing $m$ video segments consecutively:

$$P(f_{RC})^m < threshold \quad (6)$$

The value of *threshold* represents the probability of a service disruption. In order to ensure a seamless playback for a long time, a low value of *threshold* is needed. For a given *threshold*, buffer starvation is avoided when the buffer is long enough to keep playing the video when $m$ consecutive video segments are lost in the multicast channel.

In order to calculate the minimum buffer level needed to avoid buffer starvation, Fig. 3 depicts the case of losing $m = 3$ video segments consecutively. Fig. 3 shows a FLUTE server sending three video segments over MBMS. However, these video segments are lost, i.e., the AL-FEC process at the FLUTE client is not able to recover the video segments due to packet errors. The DASH client requesting the video segments does not find them on the cache, so an HTTP connection is used to retrieve the lost video segments. It is assumed that a persistent HTTP connection, previously established, is used to recover the video segments. Otherwise, an additional delay due to the TCP three-way handshake would need to be added to the analysis.

Fig. 3 shows the case in which the transmission delay for each video segment is longer than the duration of the video segment, which is the worst-case scenario in regards to buffer starvation. Since the delay accumulated for the last segment is the longest one, the buffer level must be set accordingly. Therefore, the buffer level can be calculated as:

$$d_b = trec_{(n+2)} - treq_{(n+2)} = RTT + 3D_T - 2T_{Seg} \quad (7)$$

where $treq_{(n+2)}$ is the time when the segment is requested, $trec_{(n+2)}$ is the time when the segment is received, $D_T$ is the transmission delay, and $T_{Seg}$ is the duration of the video segment, which is the same as the interval between requests.

In general, for $m$ video segments lost consecutively, when the transmission delay of the video segments is longer than the duration of the video segments ($D_T > T_{Seg}$), the minimum buffer level can be calculated as:

$$d_b = RTT + mD_T - (m-1)T_{Seg} \quad (8)$$

If $D_T < T_{Seg}$, the buffer is only needed to protect the playback during the time a unicast recovery is performed:

$$d_b = RTT + D_T \quad (9)$$

To sum up, it is possible to calculate the probability of losing a DASH segment during its transmission over eMBMS (3). This probability is calculated for a given PER and by using a mathematical model that allows simulating a Raptor code (4). This probability is then used to calculate how many segments can be lost consecutively given a *threshold* (6), which represents the probability of service disruption. Finally, the minimum buffer level needed to avoid service disruptions (8-9) is calculated taking into account the transmission delay that is introduced due to the HTTP recovery of the DASH segments. This buffer level can then be notified to the DASH clients through the *minBufferTime* field of the MPD, which is used by the DASH player to store enough data in the buffer before starting video playback.

### B. Evaluation of LTE Multimedia Broadcast Service Buffer Level

As explained above, the buffer level in an LTE Multimedia Broadcast Service depends on the probability of losing the DASH segments sent over multicast. But this probability depends on several parameters that affect the packet error rate for the LTE eMBMS deployment, as well as on the AL-FEC code rate of the multimedia broadcast service. Table II shows a set of values used to evaluate the buffer level, focusing on the evaluation of how different AL-FEC code rates affect the buffering requirements. Other parameters affecting the packet error rate have a fixed value, such as the size of the MBSFN area, the Modulation and Coding Scheme (MCS), or the Inter-Site Distance (ISD). An MBSFN area size with more cells cooperating to transmit the same signal would provide a higher SINR and therefore reduce the PER. The value selected for the MCS (15) defines a specific modulation and PHY-FEC that is robust against errors, allowing a bitrate over 1 Mbps. Finally, regarding ISD, a standard distance for the separation between base stations for urban environments has been considered. Larger values of ISD would degrade the SINR and increase the PER.

TABLE II
PARAMETERS USED TO EVALUATE AN LTE MULTIMEDIA BROADCAST SERVICE

| Parameter | Value |
|---|---|
| Coverage | 90% |
| Total number of terminals | 399 |
| Threshold | 0.00001 |
| Duration of the DASH segment | 2 s |
| Inter-Site Distance (ISD) | 500 m |
| Modulation and Coding Scheme (MCS) | 15 |
| MBSFN area size | 7 cells |
| AL-FEC code rate | 0.84 / 0.86 / 0.88 / 0.9 |

Table II also shows a fixed coverage of 90%, and a total number of terminals of 399. As it is explained below, the coverage will set a boundary for the worst case user in terms of multimedia segment transmission errors. Regarding the number of terminals, the number is selected to have a large enough population for the simulations.

The evaluation has been carried out using DASH segments of two seconds and a *threshold* (Section IV) of 0.0001, which is equivalent to a seamless video playback of 55 hours.

Using the above parameters, the probability of losing a video segment for any terminal in an MBSFN area has been calculated. These calculations combine LTE transmission simulations using OPNET[1] and a mathematical model implemented in MATLAB[2]. This model, described in [12], has been extended to calculate the minimum buffer level needed to avoid service disruptions as described in this section.

Fig. 4 shows the minimum buffer level (measured in number of DASH video segments) in relation to different probabilities of losing a video segment and for different unicast transmission delays, considering a value of RTT = 0. Transmission delays are represented as multiples of the segment duration $T_{Seg}$. Only the cases in which the loss of several consecutive segments results in an increase of the delay are represented, i.e., when $D_T > T_{Seg}$.

With the aim of providing a good coverage, the buffer level is calculated so that it is sufficient to ensure a seamless video playback for 90 percent of the users. In order to do this, the probability of losing a segment for the 10-percentile "worst" user is calculated.

---

[1] "Riverbed techonologies," [Online]. Available: http://tiny.cc/OPNET
[2] "Matlab," [Online]. Available: http://mathworks.com/products/matlab

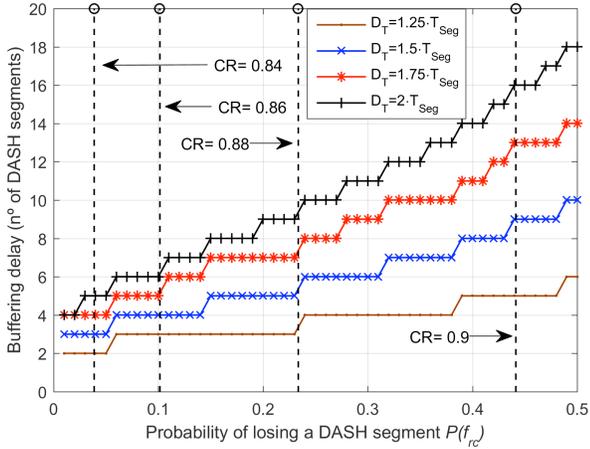

Fig. 4. Minimum buffer level that avoids service disruptions. The vertical dashed lines show the probability of the FEC decoder not recovering a segment $P(f_{RC})$ for different deployments of an LTE multimedia broadcast service.

This probability is represented in Fig. 4 in vertical dashed lines for four cases, in order of the AL-FEC code rate. For example, when using an AL-FEC code rate of 0.84, the probability of losing a video segment for the 10-percentile user is 0.0387, so the buffer level needed to avoid service disruptions would be 2, 3, 4, and 5 video segments for transmission delays of $1.25 \cdot T_{Seg}$, $1.5 \cdot T_{Seg}$, $1.75 \cdot T_{Seg}$, and $2 \cdot T_{Seg}$ respectively. For larger code rates, the probability of the FEC decoder not recovering an object is higher, and hence, a longer buffer is needed.

## V. LOW LATENCY LIVE VIDEO STREAMING

The previous section explains how to reduce the buffering delay by calculating the minimum buffer level to avoid service disruptions. Since the buffering delay depends on the duration of the segment, it could be possible to reduce the latency by reducing the duration of the segment. Wei *et al.* [13] discuss that reducing the duration of the DASH segments implies an increase in the number of HTTP requests and responses, and hence, an additional overload on the clients, the server and the network infrastructure. In an LTE multimedia broadcast service, the viability of this solution can be reconsidered, since HTTP requests are only used as an error recovery mechanism. This section explores if it is possible to further reduce latency by selecting an adequate duration for the multimedia segments and the impact of this change on the quality of the multimedia service.

### A. Selection of the DASH Segment Duration

In an LTE broadcast deployment with a good coverage, most video segments are delivered without the need for unicast recovery [12]. However, the combination of a fixed bandwidth in eMBMS with the use of AL-FEC results in a trade-off between segment size, error rate, and maximum service data rate. The maximum Service Data Rate (SDR) can be calculated as:

$$SDR = R_{eMBMS} \cdot CR \qquad (10)$$

where $R_{eMBMS}$ is the maximum bandwidth of the eMBMS channel. For a fixed error rate, reducing the video segment size would reduce the maximum service data rate, because more redundancy is needed for smaller AL-FEC source blocks. For the same reason, for a fixed service rate, reducing the video segment size would increase the error rate, and thus the number of HTTP retransmissions.

In order to look for an appropriate value of the segment size that allows reducing the delay without compromising the number of retransmissions and the quality of the video, two aspects need to be analyzed:
- The relation between segment size and AL-FEC decoder performance.
- The impact of reducing the segment size on the minimum buffer level needed to avoid service disruptions.

Fig. 5 shows the relation between segment duration and AL-FEC decoder performance, characterizing the AL-FEC performance as the number of segments that need to be recovered by HTTP. Fig. 5 can be used to select the AL-FEC providing the better service data rate for a specific limit on the percentage of segments recovered by HTTP. By mapping the 90% coverage target to a limit of 10% of the segments recovered by HTTP, Fig. 5 shows the maximum AL-FEC code rate for different segment durations: CR=0.90 for two second segments ($MCR_2$), CR=0.86 for one second segments ($MCR_1$), and CR=0.78 for 500 ms segments ($MCR_{0.5}$).

On the other hand, a minimum service data rate of 1 Mbps is recommended to provide a good compromise between video quality and bandwidth for an LTE video broadcast service [14]. This means that the minimum possible AL-FEC code rate is 0.78, which is equivalent to a service data rate of 1 Mbps as it can be calculated by using (10). So the combination of 0.5 s segments with an AL-FEC code rate of 0.78 ($MCR_{0.5}$) can be selected as a good candidate to reduce the buffering delay while keeping both the service data rate and the percentage of segments recovered by HTTP within the acceptable values for the service.

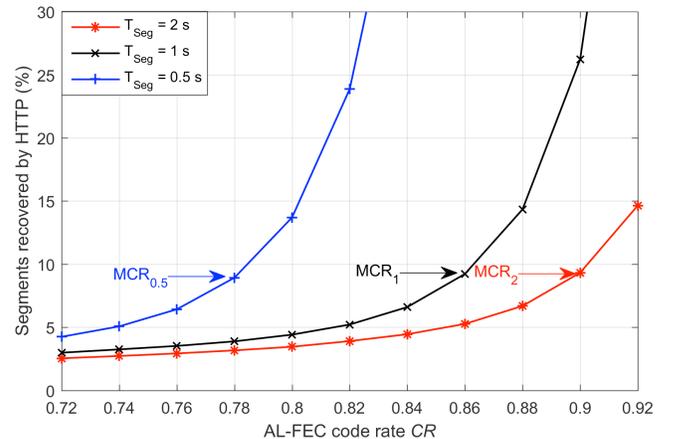

Fig. 5. Relation between segment duration, AL-FEC code rate and AL-FEC decoder performance. By taking into account a limit of 10% of the segments recovered by HTTP, the maximum AL-FEC code rates for different segment durations have been labeled as $MCR_{0.5}$, $MCR_1$, and $MCR_2$.

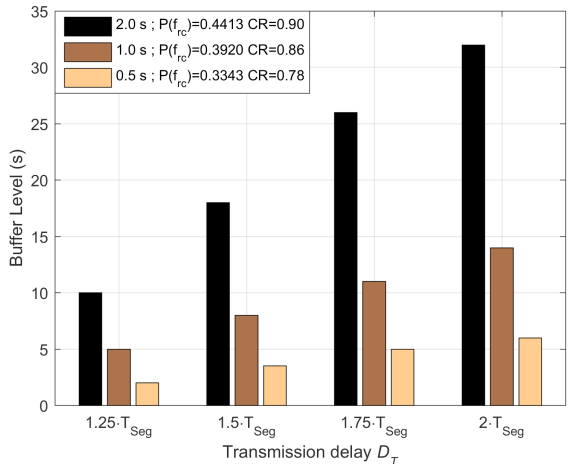

Fig. 6. Minimum buffer level needed to avoid service disruptions.

Further reducing the segment duration to 250 ms lowers the capability of the AL-FEC correcting errors, resulting in combinations of service data rate vs. segments recovered by HTTP that are not acceptable for the service. For example, an AL-FEC code rate of 0.55 implies a service data rate of 0.69 Mbps and a percentage of segments recovered by HTTP of 13.39%.

In order to analyze the impact of reducing the segment size on the buffer level, Fig. 6 shows the buffer level (in seconds) for the code rates and segment durations identified above. As shown in Section IV, different values for the transmission delay of a video segment over HTTP affect the minimum buffer level. Once again, the buffer level is calculated according to the 10-percentile user, so that 90% of the users do not perceive service disruptions.

Fig. 6 shows that for longer transmission delays the buffer level needed to avoid service disruptions must also be higher. For instance, a transmission delay of $1.25 \cdot T_{Seg}$ results in a buffer delay of 6 seconds for 2 s segments and an AL-FEC code rate of 0.9. A buffer level of 12 s would be needed for a unicast transmission delay of $2 \cdot T_{Seg}$ in the same case. On the other hand, Fig. 6 shows how using shorter duration video segments reduce the buffer requirements. Taking the $MCR_{0.5}$ combination of segment duration and AL-FEC code rate from Fig. 5 (0.5 s segments and AL-FEC code rate of 0.78), Fig. 6 shows the minimum buffer level that can be obtained for the different cases analyzed.

As a conclusion, we have shown how the results of Section IV can be applied to a case study to calculate a minimum buffer level that avoids service disruptions with a minimum latency. By notifying this value to the multimedia players using the *minBufferTime* field of the MPD this result can be used to reduce the service latency, thus having a direct impact on parameters such as the channel zapping time [15].

### B. Evaluation of the Multimedia Quality

In order to understand the relation between buffer level, service disruptions and multimedia quality in an LTE multimedia broadcast service, two different approaches have been followed to take into account perceptual video quality metrics. The evaluation is carried out considering a scenario with a multimedia segment duration of 0.5 s, a unicast transmission delay of $2 \cdot T_{Seg}$ and a service characterized by three different buffer levels: 2, 4, and 6 seconds.

The first approach is to evaluate the Peak Signal-to-Noise Ratio (PSNR) of the received video. PSNR, typically used for evaluating performance improvements of objective perceptual video quality metrics, focuses on a pixel-wise comparison of the original frame and the frame received by the multimedia player. If there are no service interruptions, the PSNR is only influenced by the encoding format. But it is not clear how PSNR can be used to measure the impact of service interruptions on the perceived quality. *Kumar and Oyman* [16] calculate the PSNR taking into account that clients try to conceal a lost frame by repeating the last successfully received frame. By adapting this approach to the LTE multimedia broadcast service, it is possible to evaluate a worst-case scenario. In this scenario, segments that cannot be retrieved in time due to a short buffer, are not played at all. Instead, the player shows the last frame of the previous successfully received segment.

The code published by the ITU-T [17] to calculate the PSNR has been combined with the LTE transmission simulation results and the mathematical model used for calculating the minimum buffer level. The calculation only considers the Luminance component (Y-PSNR), which is the most popular metric to calculate the PSNR since the visual perception is more sensitive to brightness [18].

Fig. 7 shows the resulting Y-PSNR as a function of the user percentile. Fig. 7 shows how the 10-percentile user perceives a seamless playback if the buffer level is 6 s. This is consistent with Fig. 6, which showed that it is possible to avoid any service disruptions (considering the 10-percentile user) if the segment duration is 0.5 s and the buffer level is longer than 6 s. However, for buffer levels under 6 s, service disruptions lead to loss in video quality.

In a second approach to evaluate the multimedia quality, the impact of the buffer level on the number of stalling events is analyzed for the same scenario. The number of stalling events is an objective parameter typically used as an indicator of the perceptual video quality degradation [9], [19], [20]. Based on the results of Hoßfeld *et al.* [20], we consider that the perceived quality is severely degraded when there are three or more stalling events.

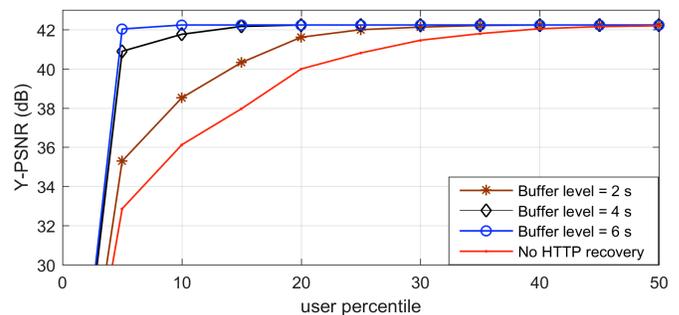

Fig. 7. Y-PSNR as a function of user percentile in case of a transmission delay of $2 \cdot T_{Seg}$.

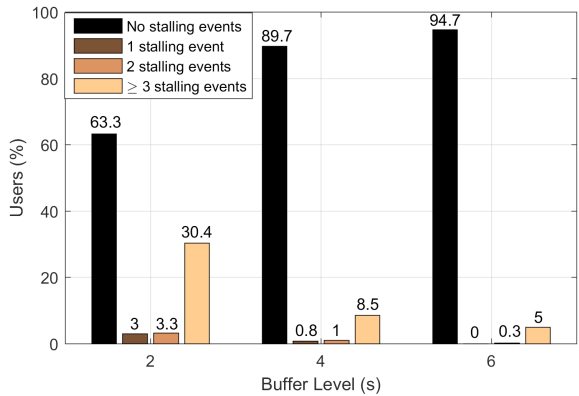

Fig. 8. Percentage of users being affected by stalling events for different buffer levels: 2 s (a), 4 s (b), and 6 s (c). The transmission delay is 2·$T_{Seg}$.

Fig. 8 shows the relation between buffer level and number of stalling events. A buffer level of 2 s is too short, so 30.4% of the users suffer from a severe video quality degradation, with three or more stalling events. A buffer level of 4 s provides an improvement but there are still 8.5% of the users with three or more stalling events. Finally, a minimum buffer level of 6 s, calculated in Section V-A for this scenario, is good to avoid quality degradation for almost 95% of the users, as shown in Fig. 8.

In conclusion, we have presented a methodology that can be used to calculate the minimum buffer level needed to avoid service disruptions in LTE multimedia broadcast services. By evaluating the perceptual video quality for different buffers we have confirmed the results obtained, comparing the quality obtained with the calculated buffer level to the quality obtained when using shorter buffers.

## VI. RELATED WORK

The provision of services consisting of the delivery of live multimedia content to multiple users poses several challenges such as how to manage the network resources efficiently or how to deliver the multimedia content with a minimum delay.

The solution to bring down the delay depends on the technology used. For instance, Yang *et al.* [21] proposed a framework to minimize the delay when an IPTV user changes the channel. The framework was based on the following idea: it is possible to reduce the delay at the price of extra download bandwidth. However, the impact of the packet error rate on the delay was not analyzed. In this paper, this problem is addressed considering the standard solution for LTE, in which video segments are recovered using both AL-FEC techniques and HTTP.

Regarding the analysis of live streaming services, several works presented analytical models to characterize delay-sensitive streaming services [22]-[24]. These models, in spite of presenting solutions to determine the minimum buffer level that avoids buffer underruns, or the current frame rate needed to reduce the delay, did not consider the use of DASH technologies for the delivery of the multimedia content.

Lohmar *et al.* [7] presented a delay analysis of a live streaming service over HTTP. The analysis determined the instant on which DASH segments are available to be requested by clients. In an LTE multimedia broadcast service, the delivery of a multimedia segment is delayed by the use of AL-FEC and unicast recovery techniques, so additional delays need to be considered to ensure the availability of DASH segments, as it is discussed in this paper.

Monserrat *et al.* [10] and De Fez *et al*. [25], [26] analyzed the time taken to deliver multimedia content over eMBMS. The analysis carried out by Monserrat *et al.* considered that video fragments lost in the multicast channel are recovered using a point-to-point connection once the transmission over eMBMS has finished. De Fez *et al*. [25], [26] defined how to use AL-FEC techniques to reduce the download time of a multimedia content that is sent using the FLUTE protocol. In this case, the proposals relied on the use of file carousels. The above recovery mechanisms are not adequate for LTE multimedia broadcast services that are supported by the architecture proposed the 3GPP. They are more appropriate for a broadcast service that is used to play a video content after it has been completely downloaded.

In the past years, several solutions addressing the problem of service disruptions for DASH have been presented [9], focusing on the adaptation algorithm of DASH and the availability of different representations coded with different bitrates. However, in an LTE video broadcast service, a single video representation is sent over the eMBMS channel, so the adaptation algorithm is not used. Section IV discusses how to calculate the minimum buffer level that avoids service disruptions, taking into account the use of AL-FEC and HTTP to recover lost segments.

A possible solution to further reduce the delay consists of the reduction of the duration of video segments. For a live streaming service over HTTP, Wei *et al.* [13] argued that this solution can increase the communication overhead. In an LTE multimedia broadcast service, video segments are sent over the eMBMS channel, so the impact of the communication overhead is significantly reduced. Section V discusses how to calculate the adequate video segment duration for an LTE multimedia broadcast service.

## VII. CONCLUSIONS

The current standard for multimedia broadcast services over LTE mobile networks integrates in a novel way elements borrowed from different standards and proposals, such as DASH to split the multimedia content in different segments, the eMBMS channel to broadcast the segments to the mobile devices using the FLUTE file delivery protocol, AL-FEC techniques to recover from packet transmission errors, and DASH (HTTP) retrieval for segments that cannot be recovered using the AL-FEC.

The combination of all these elements affect the latency of the access to multimedia content in a way that has not been studied before, though it impacts aspects of the service that are common to other broadcast technologies, such as the initial delay to start playing the content or the zapping time.

This paper has analyzed the different components of the latency, and has proposed a solution to allow a multimedia player running on a mobile device to play the multimedia content with a minimum latency. The key elements of the solution are a methodology for the calculation of an adequate value for the *availabilityStartTime* parameter included in the DASH MPD, indicating the earliest instant a segment can be accessed, and a methodology for the calculation of the adequate segment duration and the minimum buffer required for a seamless multimedia stream reproduction.

The impact of this solution on the reduction of the latency can be significant, because normally the DASH components of a multimedia streaming service are configured for the retrieval of content over a unicast TCP connection. However, the impact on the design of mobile devices is minimal, requiring only the adequate definition of parameters on the MPD used to define the multimedia content that is broadcast over LTE.

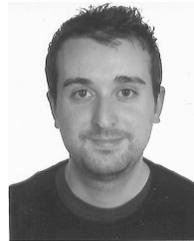

**Carlos M. Lentisco** received a Telecommunications Engineering degree (2013) from Universidad Carlos III de Madrid and a M.S. degree (2014) in Networks and Telematic Services Engineering from Universidad Politécnica de Madrid (UPM). Currently, he is a Ph.D. candidate in the Telematic Engineering Department, UPM. His research interests include multimedia, wireless networks, and quality of experience.

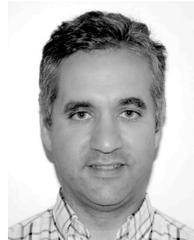

**Luis Bellido** holds a Telecommunications engineering degree (1994) from the Universidad Politécnica de Madrid (UPM), Spain, and a Ph.D. in Telecommunications Engineering (2004), also from UPM. Since 2008 he is an Associate Professor at UPM, specializing in the fields of computer networking, Internet technologies and quality of service. His current research interests include mobile networks, multimedia applications and virtualization.

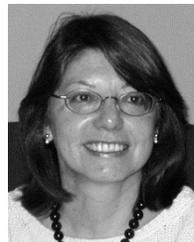

**Encarna Pastor** received the M.S. and Ph.D. degrees on Computer Science from Universidad Politécnica de Madrid (UPM) in 1980 and 1988, respectively. She is currently a Full Professor at UPM, specializing in the fields of computer networking, multimedia applications, and Internet technologies. Her current research interests include content delivery networks, multimedia networking and quality of experience.